\documentclass[pre,aps,showpacs,superscriptaddress]{revtex4}

\def \beq{\begin{equation}}         \def \eeq{\end{equation}}
\def \beqa{\begin{eqnarray}}        \def \eeqa{\end{eqnarray}}
\def \bea{\begin{array}}        \def \eea{\end{array}}

\usepackage{amsmath}
\usepackage{epsfig}
\usepackage[dvips]{color}

\begin{document}

\title{Linear response theory and transient fluctuation theorems for diffusion processes:
\\ a backward point of view}
\author{Fei Liu}
\email[Email address:]{liufei@tsinghua.edu.cn} \affiliation{Center
for Advanced Study, Tsinghua University, Beijing, 100084, China}
\author{Zhong-can Ou-Yang}
\affiliation{Center for Advanced Study, Tsinghua University,
Beijing, 100084, China} \affiliation{Institute of Theoretical
Physics, The Chinese Academy of Sciences, P.O.Box 2735 Beijing
100080, China}
\date{\today}

\begin{abstract}
{On the basis of perturbed Kolmogorov backward equations and path
integral representation, we unify the derivations of the linear
response theory and transient fluctuation theorems for continuous
diffusion processes from a backward point of view. We find that a
variety of transient fluctuation theorems could be interpreted as
a consequence of a generalized Chapman-Kolmogorov equation, which
intrinsically arises from the Markovian characteristic of
diffusion processes. }
\end{abstract}
\pacs{05.70.Ln, 02.50.Ey, 87.10.Mn} \maketitle

\section{Introduction}
One of important developments in nonequilibrium statistic physics
in the past two decades is the discovery of a variety of
fluctuation theorems (FTs) or fluctuation
relations~\cite{Evans,Gallavotti,Kurchan,Lebowitz,Bochkov77,
JarzynskiPRE97,JarzynskiPRL97,Crooks99,Crooks00,Maes,
SeifertPRL05,HatanoSasa}. These theorems were usually expressed as
exact equalities about statistics of entropy production or
dissipated work in dissipated systems. In near-equilibrium region,
these FTs reduce the fluctuation-dissipation theorems
(FDTs)~\cite{Callen,Kubo}. Hence they are also regarded as
nonperturbative extensions of  the FDTs in far-from equilibrium
region~\cite{Evans,GallavottiPRL96,Lebowitz}. Analogous to many
new findings in physics, the mathematic techniques for proofing
these theorems have been present for many decades. For instance,
thanks to the works of Lebowitz and Sphon~\cite{Lebowitz}, and
Hummer and Szabo~\cite{Hummer01}, we know that, in Markovian
stochastic dynamics these FTs have an very intimate connection
with the Kolmogorov backward equation (1931) and the applications
of the famous Feynman-Kac formula~\cite{Feynman,Kac} (1948) and
Girsanov formula~\cite{Cameron,Girsanov} (1960). The involvement
of the backward equation or more precisely, its perturbed versions
in deriving the FTs is not occasional. Previous many works have
proved that various FTs originate from the symmetry-breaking of
time reversal in dissipated
systems~\cite{Evans,Gallavotti,Kurchan,Lebowitz,Bochkov77,
JarzynskiPRE97,JarzynskiPRL97,Crooks99,Crooks00,Maes,
SeifertPRL05,HatanoSasa,Kurchanrev}. This point is now widely
accepted and reader may reference an excellent synthesis from this
point of view by Chetrite and Gawedzki~\cite{Chetrite08CMP}.
Intriguingly, the backward equation concerns about, at future time
given a state or a subset, how system evolves in it from a past
time. Namely, the backward equation is a final value problem, and
can be evaluated backward in time from future to past. Hence, the
backward rather than the forward equation or Fokker-Planck
equation is natural tool to describe time reversal. Actually, this
idea has been implied earlier in finding conditions for the
detailed balance principle of homogeneous Markov stochastic
systems~\cite{Risken,Gardiner}. In this work, we roughly call a
discussion on the basis of past time \emph{backward} to
distinguish more conventional discussion on the basis of future
time (forward).

Although thes FTs are of importance and extensive attention was
paid on them in past two decade, there were fewer works concerning
about this connection during a long time. The reasons may be two
sides. On one hand, physicists are not very familiar with the
backward equation compared with Fokker-Plank equation.
Introduction about the backward equation in many classic
books~\cite{Risken,Gardiner} was usually about its equivalence
with forward equation. Its application is solely first passage
time or exit problems. On the other hand, as mentioned perviously,
time reversal is very relevant to the FTs. Most of theorems could
be evaluated by the ratio of probability densities of observing a
stochastic trajectory and its reverse in a stochastic system and
its time reversal,
respectively~\cite{Chernyak,Kurchan,Crooks00,Crooks99}. Hence
physicists familiar with quantum physics may favor the direct path
integral approach~\cite{Onsager,ZinnJustin}. Until recently, some
works began to investigate and exploit the connection between the
FTs and the backward equation~\cite{Chetrite08CMP,Ge,LiuF1,LiuF2}.
For instance, Ge and Jiang~\cite{Ge} employed a perturbed backward
equation and Feynman-Kac formula to reinvestigate Hummer and
Szabao's earlier derivation~\cite{Hummer01} about the celebrated
Jarzynski equality~\cite{JarzynskiPRE97,JarzynskiPRL97} from
mathematical rigors. A generalized multidimensional version of the
equality was obtained. On the basis of an abstract time reversal
argument, Chetrite and Gawedzki~\cite{Chetrite08CMP} established
an exact fluctuation relation between the perturbed Markovian
generator of forward process and the generator of time-reversed
process, though the authors did not use perturbed backward
equations explicitly. Inspired by Ge and Jiang's idea, we obtained
two time-invariable integral identities for very general discrete
jump and diffusion process, respectively~\cite{LiuF1,LiuF2}.
Considering that several transient integral fluctuation
theorems~\cite{JarzynskiPRE97,JarzynskiPRL97,Maes,SeifertPRL05,HatanoSasa}
are their path integral representations in specific cases, we
called these two integral identities generalized integral
fluctuation theorems (GIFTs). Our further analysis showed that
these GIFTs had well-defined time reversal explanations that are
consistent with those achieved by Chetrite and
Gawedzki~\cite{Chetrite08CMP}. Hence, their detailed versions or
the transient detailed fluctuation theorems (DFTs) should be
easily established. In addition to simplicity in evaluations, to
us the most impressive point of using perturbed backward equations
is that a specific time reversal is defined naturally and
explicitly given a specific IFT, and the latter can be designed
``freely" from the GIFTs. This apparently contrasts with
conventional direct path integral approach (including
Ref.~\cite{Chetrite08CMP}), which requires a specific time
reversal first and then obtains a specific IFT. Previous works
showed the definition of time reversal may be nontrivial, e.g.
that in Hatano-Sasa equality~\cite{HatanoSasa}.

The aims of this work are two-fold. First, we attempt to present a
comprehensive version of our previous work about continuous
diffusion process~\cite{LiuF1}. In addition that many details that
were missed or very briefly reported previously will be made up,
which mainly includes classification of the existing IFTs and time
reversals and derivation of the transient DFTs from a point of
view of the GIFT, we also present several new theoretical results.
The most significant progress is to find that the time-invariable
integral identity we obtained previously is a generalized
Chapman-Kolmogorov equation in general diffusion processes; the
path integral representation of the well-known Chapman-Kolmogorov
equation may be regarded as the first IFT. Additionally, we
uniformly obtain the GIFT for the Smoluchowski~\cite{Smoluchowski}
and Kramers type~\cite{Kramers} diffusions by employing a limited
Girsanov formula (see Appendix~\ref{extendedCMGformula}). In
previous works~\cite{Lebowitz,Chetrite08CMP} the latter was
considered individually. Our second aim is to show that there is
an alternative way using the backward equation to derive the
classical linear response theory~\cite{Callen,Kubo}, and a simple
extension of this ``lost" approach results into the transient FTs
found almost forty years later. Although it is widely accepted
that the FTs reduce to the linear response theory when they are
approximated linearly near
equilibrium~\cite{Evans,GallavottiPRL96,Lebowitz,Chetrite08CMP},
one may see a significant difference between their derivations: in
books~\cite{Risken} the linear response theory always starts from
an evaluation of probability distribution function using
time-dependent perturbation theory, whereas the former did not use
this function at all. We show this differences may be obviously
diminished if one employs the backward equation to evaluate the
linear responses of perturbed systems at the very beginning.
Moreover, this reevaluation evokes our attention to the importance
of the Chapman-Kolmogorov equation. We are tempting to think
whether the dominated forward idea using the forward equation
postpones the findings of the transient FTs in Markovian
stochastic dynamics.

The organization of this work is as following. We first present
some essential elements about the continuous diffusion process in
sec.~\ref{reviewofstochastictheory}. The Chapman-Kolmogorov
equation, Feynman-Kac and Girsanov formulas are explained. In
sec.~\ref{linearresponsetheory}, we derive the linear response
theory using the backward equation. Two FDTs that recently
attracted considerable interest are also discussed briefly.
Section~\ref{GeneralizedIntegralfluctuationtheorem} mainly devotes
the GIFT, which includes the relationship between the GIFT and the
generalized Chapman-Kolmogorov equation, time reversal explanation
of the GIFT, and classification of the IFTs and time reversals in
the literature from a point of view of the GIFT. Additionally, we
also propose a Girsanov equality and explain differences between
this equality and the GIFT. In
sec.~\ref{GeneralizedCrooksrelation}, we derive the detailed
version of the GIFT on the basis of its the time reversal
explanation. We summarize our conclusions in
sec.~\ref{Conclusion}.

\section{Elements of stochastic diffusion process}\label{reviewofstochastictheory}
We consider a general $N$-dimension stochastic system ${\bf
x}$$=$$\{x_i\}$, $i$$=$$1$, $\cdots$, $N$ described by a
stochastic differential equation (SDE)~\cite{Gardiner}
\begin{eqnarray}
d{\bf x}(t)={\bf A}({\bf x},t)dt + {\bf B}^{\frac{1}{2}}({\bf
x},t)d{\bf W}(t), \label{SDE}
\end{eqnarray}
where $d{\bf W}$ is an $N$-dimensional Wiener process, ${\bf
A}$$=$$\{A_i\}$ denotes a $N$-dimensional drift vector, and ${\bf
B}^{1/2}$ is the square root of a $N$$\times$$N$ semipositive
definite and symmetric diffusion matrix ${\bf B}$
\begin{eqnarray}
\label{diffusionmatrix}
 \bf{B}=\left[
  \begin{array}{cc}
   \bf{D}& 0 \\
    0 & 0 \\
  \end{array}
\right],
\end{eqnarray}
where $\bf D$ is a $M$$\times$$M$ ($M$$\le$$N$) positive definite
submatrix. We call a stochastic process Smoluchowski
(nondegenerate) type for $M$$=$$N$, and Kramers (degenerate) type
otherwise, because the Smoluchowski and Kramers
equations~\cite{Smoluchowski,Kramers} are their typical examples.
One usually converts the SDE into two equivalent partial
differential equations of transition probability density
$\rho({\bf x},t|{\bf x}',t')$ ($t>t'$): the forward or
Fokker-Planck equation
\begin{eqnarray}
\label{FPoperator}
\partial_t \rho={\cal L}({\bf x},t)\rho=[-\partial_{x_i} A_i({\bf x},t)
+\frac{1}{2}\partial_{x_i}\partial_{x_l} B_{il}({\bf x},t)]\rho,
\end{eqnarray}
and the Kolmogorov backward equation
\begin{eqnarray}
\label{Kbackoperator}
\partial_{t'} \rho=-{\cal L}^+({\bf x}',t') \rho=-[A_i({\bf x}',t)\partial_{x_i'} + \frac{1}{2}B_{il}({\bf
x}',t)\partial_{x_i'}\partial_{x_l'}]\rho.
\end{eqnarray}
The initial and final conditions of them are $\delta({\bf x}-{\bf
x}')$, respectively. We follow Ito's convention for the SDE and
use Einstein's summation convention throughout this work unless
explicitly stated. The forward equation defines a probability
current ${\bf J}[\rho({\bf x},t)]$, components of which are
\begin{eqnarray}
\label{probcurrent} J_i[\rho({\bf x},t)]= A_i({\bf
x},t)\rho-\frac{1}{2}\partial_{x_l}[B_{il}({\bf x},t)\rho],
\hspace{0.3cm}{\rm and}\hspace{0.3cm} {\cal L}({\bf x},t)\rho({\bf
x},t)=-\partial_{x_i}J_i[\rho({\bf x},t)].
\end{eqnarray}
Different from the forward equation, Eq.~(\ref{Kbackoperator}) is
about past time $t'$, and generally $\rho({\bf x},t|{\bf x}',t')$
does not have a probability interpretation with respect to
variable ${\bf x}'$. The connection between the forward and
backward equations may be seen from the famous Chapman-Kolmogorov
equation~\cite{Gardiner}
\begin{eqnarray}
\label{ChapmanKolmogorov} \rho({\bf x}_2,t_2|{\bf x}_1,t_1)=\int
d{\bf x}\rho({\bf x}_2,t_2|{\bf x},t)\rho({\bf x},t|{\bf
x}_1,t_1),
\end{eqnarray}
where $t_1$$\le$$t$$\le$$t_2$. An equivalent expression is its
derivative with respect to time $t$,
\begin{eqnarray}
\label{differChapmanKolmogorov} 0&=& \partial_{t}[\int d{\bf
x}\rho({\bf
x}_2,t_2|{\bf x},t)\rho({\bf x},t|{\bf x}_1,t_1)]\nonumber\\
&=&\int d{\bf x}[\partial_t\rho({\bf x}_2,t_2|{\bf x},t)]\rho({\bf
x},t|{\bf x}_1,t_1)+\rho({\bf x}_2,t_2|{\bf
x},t)[\partial_t\rho({\bf x},t|{\bf x}_1,t_1)].
\end{eqnarray}
The reason of the left hand side vanishing is very obvious.
Equation~(\ref{differChapmanKolmogorov}) implies the operators
${\cal L}$ and ${\cal L}^{+}$ are adjoint each other if one
substitutes the time-derivatives on the right hand side with
forward and backward equations. Conversely, through the same
equation we can as well obtain the backward (forward) equation
using the adjoint characteristic of the operators if known the
forward (backward) equation first.

There are two famous formulas in stochastic theory that are
employed in this work. One is the Feynman-Kac formula, which was
originally found by Feynman in quantum mechanics~\cite{Feynman}
and extended by Kac~\cite{Kac} in stochastic process. Assuming a
partial differential equation
\begin{eqnarray}
\label{FKpartialdiffequation}
\partial_{t'} u({\bf x},t')=-{\cal L}^+({\bf x}',t')u({\bf x},t')-g({\bf x},t')u({\bf
x},t'),
\end{eqnarray}
with a final condition $u({\bf x},t)=q({\bf x})$, then its
solution has a path integral representation given by
\begin{eqnarray}
\label{FKformula} u({\bf x},t')={}^{{\bf
x},t'}\langle\exp[\int^t_{t'} g({\bf x}(\tau),\tau)d\tau]q[{\bf
x}(t)] \rangle.
\end{eqnarray}
where the expectation ${}^{{\bf x},t'}\langle$ $\rangle$ is an
average over all trajectories $\{{\bf x}(\tau)\}$ determined by
SDE~(\ref{SDE}) taken conditioned on ${{\bf x}(t')={\bf x}}$.
Letting $g$$=$$0$ and $q({\bf x})$ be a $\delta$-function, the
Feynman-Kac formula also gives a path integral representation of
backward equation~(\ref{Kbackoperator}). The other is the Girsanov
formula~\cite{Girsanov}. Roughly speaking, the standard version of
this formula is about probability densities of observing the same
trajectory $\{{\bf x}(\tau)\}$ between time $t_0$ and $t$ in two
different stochastic systems: Assuming they have the same
nondegenerate diffusion matrix ${\bf B}$ [$=$${\bf D}$ in
Eq.~(\ref{diffusionmatrix})] and one of them (denoted by prime)
differs from the other only in the drift vector, $A'_i=A_i+a_i$,
then the probability densities ${\cal P}'$ and ${\cal P}$ are
related by
\begin{eqnarray}
\label{CMGratio} {{\cal P}'[\{{\bf x}(\tau)\}]} = {{\cal P}[\{{\bf
x}(\tau)\}]}e^{-\int_{t_0}^t {\cal R}[{\bf a}](\tau,{\bf
x}(\tau))d\tau}
\end{eqnarray}
and
\begin{eqnarray}
\label{CMGformula} {\cal R}[{\bf a}]=\frac{1}{2} a_i({\bf
B}^{-1})_{il}a_l -a_i({\bf B}^{-1})_{il}(v_l-A_l),
\end{eqnarray}
where $v_i$$=$$dx_i/d\tau$ and the integral is defined by Ito
stochastic integral. The inverse of the diffusion matrix above
indicates the indispensability of the nondegenerate characteristic
of these diffusions. Nevertheless, degenerate cases are more
generic in real physical models, e.g., the Kramers
equation~\cite{Kramers}. After recalled the original evaluation of
the Girsanov formula, we find a limited version specifically
aiming at the degenerate diffusions; see
Appendix~\ref{extendedCMGformula}.

\section{Linear response theory}\label{linearresponsetheory}
Evaluating linear response of a system to an external perturbation
is essential ingredient of the fluctuation-dissipation
theorems~\cite{Callen,Kubo}. For stochastic diffusion system, the
conventional approach was based on the forward Fokker-Plank
equation and applied the time-dependent perturbation
theory~\cite{Risken,Marconi}. Here we show that the same results
can be also achieved using the Kolmogorov backward equation. Our
approach is not only relatively simple, but also its theoretical
results are able to be extended to the later transient FTs
naturally.

Assuming a perturbed stochastic system having a Fokker-Planck
operator ${\cal L}_{\rm p}={\cal L}_{\rm o}({\bf x},t)+{\cal
L}_{\rm e}({\bf x},t)$, where ${\cal L}_{\rm o}$ and ${\cal
L}_{\rm e}$ are unperturbed (denoted by the subscript ``o") and
perturbed (denoted by the subscript ``p") components,
respectively, and that the perturbation is applied at time 0. For
the sake of generality, the unperturbed system may be stationary
or nonstationary, and the type of perturbation is arbitrary.
Further assuming the probability distribution functions of the
unperturbed and perturbed systems be $f_{\rm o}({\bf x},t)$ and
$f_{\rm p}({\bf x},t)$, respectively. For a physical observable
$B({\bf x})$, one may define its dynamic version $B_{\rm p}(t|{\bf
x},t')$ by
\begin{eqnarray}
\label{defdynamicobserval} B_{\rm p}(t|{\bf x},t')=\int d{{\bf
x}}'B({\bf x}')\rho_{\rm p}({\bf x}',t|{\bf x},t')={}^{{\bf
x},t'}\left\langle B({\bf x}(t))\right\rangle_{\rm p},
\end{eqnarray}
where $\rho_{\rm p}$ is the transition probability density and
$B_{\rm p}(t|{\bf x},t)=B({\bf x})$. Mean value of the observable
at time $t$ is then evaluated by
\begin{eqnarray}
\label{mean&dynamicobserval} \langle B\rangle_{\rm p}(t)=\int
d{{\bf x}}B_{\rm p}(t|{\bf x},0)f_{\rm p}({\bf x},0)=\int d{{\bf
x}}B_{\rm p}(t|{\bf x},0)f_{\rm o}({\bf x},0).
\end{eqnarray}
Obviously, the dynamic observable~(\ref{defdynamicobserval})
satisfies a backward equation analogous to
Eq.~(\ref{Kbackoperator})
\begin{eqnarray}
\label{evoleqofdynamicobs}
\partial_{t'} B_{\rm p}(t|{\bf x},t')&=&-{\cal L}^{+}_{\rm o}({\bf x},t')B_{\rm p}(t|{\bf x},t')
-{\cal L}^+_{\rm e}({\bf x},t')B_{\rm p}(t|{\bf x},t'),
\end{eqnarray}
where ${\cal L}^+_{\rm e}$ is the adjoint operator of ${\cal
L}_{\rm e}$. The Chapman-Kolmogorov
equation~(\ref{ChapmanKolmogorov}) also holds for the dynamic
observable given by
\begin{eqnarray}
\label{ChapmanKolmogorovFDcase}
\partial_{t'}[\int d{\bf x}B_{\rm
p}(t|{\bf x},t')f_{\rm p}({\bf x},t')]=0.
\end{eqnarray}
Equation.~(\ref{mean&dynamicobserval}) may be regarded as a direct
consequence of the above identity.

The linear approximation solution of
Eq.~(\ref{evoleqofdynamicobs}) may be obtained by two approaches.
The first one is to use the standard perturbation technique and to
regard the last term in the equation as a small perturbation. We
expand the dynamic observable to first order
\begin{eqnarray}
\label{linearapproximation} B_{\rm p}(t|{\bf x},t')=B_{\rm o}(t|{\bf
x},t')+B_1(t|{\bf x},t') + \cdots,
\end{eqnarray}
and impose their final conditions $B_{\rm o}(t|{\bf x},t)=B({\bf
x})$ and $B_1(t|{\bf x},t)=0$. Substituting it into
Eq.~(\ref{evoleqofdynamicobs}), we obtain the zero and first order
terms satisfying
\begin{eqnarray}
\partial_{t'} B_{\rm o}(t|{\bf x},t')&=&-{\cal L}_{\rm
o}^{+}({\bf x},t')B_{\rm o}(t|{\bf
x},t'),\nonumber\\
\partial_{t'} B_1(t|{\bf x},t')&=&-{\cal L}_{\rm
o}^{+}({\bf x},t')B_1(t|{\bf x},t')- {\cal L}^+_{\rm e}({\bf x},t')
B_{\rm o}(t|{\bf x},t'),
\end{eqnarray}
respectively, and their solutions have path integral
representations (e.g. Theorem 7.6 in Ref.~\cite{Karatzas})
\begin{eqnarray}
B_{\rm o}(t|{\bf x},t')&=&{}^{{\bf x},t'}\left\langle B({\bf
x}(t))\right\rangle_{\rm
o},\label{KaratzasRep1}\\
B_1(t|{\bf x},t')&=&{}^{{\bf x},t'}\langle \int_{t'}^t d\tau {\cal
L}^+_{\rm e}({\bf x},\tau)B_{\rm o}(t|{\bf
x}(\tau),\tau)\rangle_{\rm o},\label{KaratzasRep2}
\end{eqnarray}
respectively. $B_{\rm o}(t|{\bf x},t')$ is obviously the dynamic
observable in the unperturbed system. Then the linear
approximation of the mean of the observable is
\begin{eqnarray}
\label{linearexpansionFDR} \langle B\rangle_{\rm p}(t)&=&\langle
B\rangle_{\rm o}(t) +\int_0^td\tau \int d{\bf x}f_{\rm
o}({\bf x},\tau){\cal L}^+_{\rm e}({\bf x},\tau)B_{\rm o}(t|{\bf x},\tau)+\cdots\nonumber \\
&=&\langle B\rangle_{\rm o}(t)+\int_0^td\tau\left\langle
[f^{-1}_{\rm o}{\cal L}_{\rm e}(f_{\rm
o})](\tau)B(t)\right\rangle_{\rm o}+\cdots,
\end{eqnarray}
where $\langle$ $\rangle$ denotes the average over the
trajectories starting from initial distribution function $f_{\rm
o}({\bf x},0)$, and we used the adjoint characteristic of ${\cal
L}_{\rm e}$ in the second line. Then, we can obtain familiar
response functions by substituting concrete perturbation
expressions in the above equation. The second approach is more
direct and interesting. Let us consider a ``twisted"
Chapman-Kolmogorov equation
\begin{eqnarray}
\label{exactdynamicequation} \partial_{t'}[\int d{\bf x} B_{\rm
p}(t|{\bf x},t')f_{\rm o}({\bf x},t')]&=&-\int d{\bf x}f_{\rm
o}({\bf x},t'){\cal L}_{\rm e}^+ B_{\rm p}(t|{\bf x},t').
\end{eqnarray}
We must emphasize this is exact. Integrating both sides with
respect to time $t'$ from $0$ to $t$, we immediately see the left
hand side is just the minus of the difference between the means of
the observable in the perturbed and unperturbed systems. If the
first order approximation was concerned about, namely, the
subscript ``p" is replaced by ``o" on the right hand side of
Eq.~(\ref{exactdynamicequation}), we reobtain
Eq.~(\ref{linearexpansionFDR}). Compared with conventional
approaches on the basis of the forward equation, these two
approaches here do not need time-ordering operator or interaction
representation~\cite{Hanggi,Diezemann}. Particularly, in our
second approach we even do not need the time-dependent
perturbation theory and path integral representation.

\subsection{Fluctuation-dissipation theorems}
The classical fluctuation-dissipation theorems state that the
linear response function of an equilibrium system to a small
perturbation is proportional to the two-point time-correlation
function of the unperturbed system~\cite{Kubo,Callen}. This topic
is attracting considerable interest due to continuous efforts of
extending the standard one to nonequilibrium
region~\cite{Hanggi,Diezemann,
Baiesi,Chetrite08JSM,SpeckEurophys06}. Here we briefly discuss two
intriguing FDTs~\cite{Baiesi,Chetrite08JSM} in two typical
physical models. In addition to preparing some definitions of two
models that will be used in following sections, we want to show
that, although these two theorems are nontrivial in physical
interpretation, they may be regarded as simple applications of two
general identities
\begin{eqnarray}
\partial_{x_i}[B_{ij}(\partial_{x_j}E)f ]&=& 2[{\cal L}(Ef )-E{\cal L}(f)
+(\partial_{x_i}E)J_i(f)]\label{identity1}\\
&=&{\cal L} (Ef)-E{\cal L} (f) +{\cal L}^+(E)f,\label{identity2}
\end{eqnarray}
where $E$ and $f$ are arbitrary functions. They should be used in
previous works. Interestingly, we find these two identities are
still very useful in the transient FTs. There we will use a new
identity derived from them
\begin{eqnarray}
{\cal L}(Ef )={\cal L}^+(E)f-{\cal L}(f)E-
2\partial_{x_i}[J_i(f)E].\label{identity3}
\end{eqnarray}

\subsubsection{Overdamped Brownian motion}\label{FDRoverdampedBM}
Multidimensional overdamped Brownian motion is a typical example
of the Smoluchowski type diffusions~\cite{Lax}, the SDE equation
of which is simply
\begin{eqnarray}
\label{multidimensionaloverdampedBM} d{\bf x}={\bf M}({\bf
x},t)[-\nabla U({\bf x},t)+{\bf F}({\bf x},t)]dt+{\bf
B}^{\frac{1}{2}}({\bf x},t)d{\bf W}(t),
\end{eqnarray}
where ${\bf F}$ is a nonconservative additive force, the
nonnegative mobility and diffusion matrixes are related by $2{\bf
M}$$=$$\beta{\bf B}$, and $\beta^{-1}$$=$$k_{\rm B}T$ with
Boltzmann constant $k_{\rm B}$ and coordinate-independent
environment temperature $T$. We assume perturbation is realized by
adding a time-dependent potential $-h(t)V({\bf x})$ to the
original one $U({\bf x},t)$. Under this circumstance the perturbed
component ${\cal L}_{\rm e}$ is $-h(t')\partial_{x_i}
M_{il}(\partial_{x_l} V)$. Substituting it into
Eq.~(\ref{linearexpansionFDR}), we obtain the response function
\begin{eqnarray}
R_B(t,\tau)&=&\delta \langle B\rangle_{\rm p}(t)\left/\delta
h(\tau) \right|_{h=0}\nonumber\\
&=&\left\langle f_{\rm o}^{-1}(\tau)\partial_{x_i}[f_{\rm o}
M_{il}\partial_{x_l} V](\tau)B(t)\right\rangle_{\rm o}.
\label{responsefun}
\end{eqnarray}
This expression seems very different from the standard
FDT~\cite{Kubo}, even if the unperturbed system is in equilibrium.
However, this difference is not intrinsic. Choosing ${\cal
L}$$=$${\cal L}_{\rm o}$ the Fokker-Planck operator of
Eq.~(\ref{multidimensionaloverdampedBM}) and $E$$=$$V({\bf
x},t')$, and noticing that the left hand side of
Eq.~(\ref{identity1}) is just $2{\cal L}_e(f_0)/h(t')\beta$, we
obtain two new expressions of Eq.~(\ref{responsefun}) given
by~\cite{Chetrite08JSM,Baiesi}
\begin{eqnarray}
R_B(t,\tau)&=&\beta \frac{d}{d\tau}\langle V(\tau)B(t)\rangle_{\rm
o}- \beta\left\langle [f_{\rm o}^{-1}J_i(f_{\rm
o})\partial_{x_i}V](\tau)B(t)\right\rangle_{\rm
o}\label{ChetriteFDR}\\
&=&\frac{\beta}{2} \frac{d}{d\tau}\langle V(\tau)B(t)\rangle_{\rm
o}- \frac{\beta}{2}\left\langle {\cal L }^+_{\rm
o}(V)(\tau)B(t)\right\rangle_{\rm o}.\label{MaseFDR}
\end{eqnarray}
Although the FDT~(\ref{ChetriteFDR}) still faces the difficulty of
unknown probability distribution $f_{\rm o}$ as
Eq.~(\ref{responsefun}), it intuitively indicates that the
responses are different for the unperturbed systems prepared in
equilibrium and nonequilibrium states; the latter usually has
nonvanishing probability current. In contrast, the
FDT~(\ref{MaseFDR}) does not need this distribution and is more
useful in practical simulation or experiment. The second term on
the right hand side was interpreted as a correlation with
dynamical activity~\cite{Baiesi}.

\subsubsection{Underdamped Brownian motion}\label{FDRunderdampedBM}
The second model is one-dimension underdamped Brownian motion (no
apparent differences in discussion for multidimensional case),
\begin{eqnarray}
\label{underampedBM} dp&=&-\partial_x{{\cal H}_0(x,p,t)}
dt+F(x,t)dt-{\gamma_0}p
dt+ \sqrt{2m{\gamma_0}/\beta}\hspace{0.1cm}dW\nonumber\\
dx&=&\partial_p{{\cal H}_0(x,p,t)}dt.  \nonumber
\end{eqnarray}
The deterministic Hamiltonian system is included by choosing
$\gamma_0$$=$$0$ and $F$$=$$0$. For convenience, we rewrite this
SDE into a matrix form
\begin{eqnarray}
d{\bf r}= \Pi\cdot\nabla {H_0}dt +{\bf F} dt-\Gamma\cdot {\bf P}dt
+ \sqrt{2m\Gamma/\beta} \hspace{0.1cm}d{\bf W},
\end{eqnarray}
where we define new vectors ${\bf r}^{\rm
T}$$=$$[r_1,r_2]$$=$$[p,x]$, ${\bf P}^{\rm T}$$=$$[p,0]$, ${\bf
F}^{\rm T}$$=$$[F,0]$, $\nabla^{\rm T}$$=$$[\partial_p,
\partial_x]$, and matrixes
\begin{eqnarray}
\Pi=\left[
      \begin{array}{cc}
        0 & -1 \\
        1 & 0 \\
      \end{array}
    \right],\hspace{1cm}
\Gamma=\left[
      \begin{array}{cc}
        {\gamma_0} & 0 \\
        0 & 0 \\
      \end{array}
    \right].
\end{eqnarray}
This is a typical example of Kramers type diffusions. According to
the types of the perturbations, several different FDTs with
specific conditions may be obtained. The relatively simple case is
that the perturbation is still through a potential $-h(t)V({x})$
and ${\cal L}_{\rm e}$$=$$-h(t)(\partial_{x}V)\partial_p$. We can
of course obtain a FDT as Eq.~(\ref{responsefun}) by directly
substituting ${\cal L}_{\rm e}$ into
Eq.~(\ref{linearexpansionFDR}) (not shown here). In addition, one
may expect that the left hand side of Eq.~(\ref{identity1}) is
still proportional ${\cal L}_{\rm e}(f_{\rm o})$ as that in the
overdamped case. This is indeed true if choosing
$E$$=$$p\partial_xV(x)$ and assuming ${\gamma_{0}}$ independent of
spatial and momentum coordinates. We obtain
\begin{eqnarray}
R_B(t,\tau)&=&\frac{\beta }{{\gamma_{0}}m }
\frac{d}{d\tau}\left\langle (p\partial_x
V)(\tau)B(t)\right\rangle_{\rm o}- \frac{\beta }{{\gamma_{0}} m
}\left\langle [f_{\rm o}^{-1}J_i(f_{\rm
o})\partial_{r_i}(p\partial_xV)](\tau)B(t)\right\rangle_{\rm
o}\label{overdampChetriteFDR}\\
&=&\frac{\beta}{2{\gamma_{0}} m } \frac{d}{d\tau}\langle
(p\partial_xV)(\tau)B(t)\rangle_{\rm o}-\frac{\beta }{2{\gamma_{0}}
m }\left\langle {\cal L }^+_{\rm
o}(p\partial_xV)(\tau)B(t)\right\rangle_{\rm o}
.\label{overdampMaseFDR}
\end{eqnarray}
These new FDTs seem to be very different from
Eqs.~(\ref{ChetriteFDR}) and~(\ref{MaseFDR}) in the overdamped
case. For instance, Eq.~(\ref{overdampChetriteFDR}) is not as good
as Eq.~(\ref{ChetriteFDR}) in concept because the current
$J_{r_i}$ are not zero even if the unperturbed system has
canonical distribution [in equilibrium and ${\bf F}$$=$$0$].
Particularly, these FDTs cannot automatically reduce to the
standard FDT~\cite{Kubo} in deterministic Hamiltonian system by
simply choosing ${\gamma_{0}}$$=$$0$. These problems could be
avoided if one notices the left hand side of Eq.~(\ref{identity1})
vanishes for $E$$=$$V(x)$ (the same consequence as vanishing
${\gamma_{0}}$) and introduces a modified current
\begin{eqnarray}
\tilde{{\bf J}}(f_{\rm o})&=&{\bf J}(f_{\rm
o})+\beta^{-1}\Pi\nabla f_{\rm o}.
\end{eqnarray}
Then we obtain another FDT given by
\begin{eqnarray} R_B(t,\tau)&=&\beta
\frac{d}{d\tau}\langle V(\tau)B(t)\rangle_{\rm o}- \beta\langle
[f_{\rm o}^{-1}{\tilde J}_{r_i}(f_{\rm
o})\partial_{r_i}V](\tau)B(t)\rangle_{\rm o}.
\label{overdampedChetriteFDRVcase}
\end{eqnarray}
This expression is the same as Eq.~(\ref{ChetriteFDR}), and the
last term vanishes for an unperturbed system having canonical
distribution. We must emphasize that
Eq.~(\ref{overdampedChetriteFDRVcase}) is suitable to the cases
that $\gamma_0$ is any function of the coordinate ${\bf r}$.

For general perturbations that depend on spatial and momentum
coordinates simultaneously, e.g. $-h(t)V(x,p)$~\cite{Marconi}, the
above FDTs usually do not hold. Considering simple case that
${\gamma_{0}}$ is time-dependent only.
Equation~(\ref{overdampedChetriteFDRVcase}) is then modified by an
additional term
\begin{eqnarray}
+\beta m \gamma_0\langle \{[f_{\rm o}^{-1}{\tilde J}_{x}(f_{\rm
o})
-p/m]\partial_pV+\beta^{-1}\partial^2_pV\}(\tau)B(t)\rangle_{\rm
o}. \label{overdampedChetriteFDRVcaseMOD}
\end{eqnarray}
Finally, if we temporarily forget the time derivative in these
previous FDTs, we can obtain a more concise FDT
\begin{eqnarray}
R_B(t,\tau)&=&\beta\left\langle [A_{r_i}
\Omega_{il}\partial_{r_l}V](\tau)B(t)\right\rangle_{\rm o}-
\beta\langle [f^{-1}_{\rm o} {\tilde
J}_{r_i}\Omega_{il}\partial_{r_l}V](\tau)B(t) \rangle_{\rm o},
\label{overdampedChetriteFDRpqdep}
\end{eqnarray}
where matrix $\Omega$ is $-\Pi(m\Gamma-\Pi)^{-1}$. One may easily
prove that Eqs.~(\ref{overdampedChetriteFDRpqdep}) and
~(\ref{overdampedChetriteFDRVcase}) are identical if the potential
$V$ is a function of the spatial coordinate $x$ only or vanishing
$\gamma_0$. The above discussion about the FDTs in these two
physical model are mainly technical. Their underlying physics may
reference previous literature~\cite{Baiesi,Chetrite08JSM}.

\section{Generalized Integral fluctuation theorem }\label{GeneralizedIntegralfluctuationtheorem}
During the reinvestigation of the linear response on the basis of
the backward equation, we notice that the Chapman-Kolmogorov
equation~(\ref{ChapmanKolmogorovFDcase}) plays a role implicitly.
Particularly, we find that there are other functions or variants
of $B_{\rm p}(t|{\bf x},t')$ not only satisfying the same
Chapman-Kolmogorov equation but also having the same mean of the
observable, e.g., $B'_{\rm p}(t|{\bf x},t')$ satisfying
\begin{eqnarray}
\label{mutevoleqofdynamicobs}
\partial_{t'} B'_{\rm p}(t|{\bf x},t')&=&-{\cal L}^{+}_{\rm p}({\bf x},t')B'_{\rm p}(t|{\bf x},t')
+f^{-1}_{\rm p}({\bf x},t')\left[f_{\rm p} {\cal L}^+_{\rm e}
-{\cal L}_{\rm e}(f_{\rm p})\right]({\bf x},t') B'_{\rm p}(t|{\bf
x},t')\nonumber\\
&=&-{\cal L}^{+}_{\rm o}({\bf x},t')B'_{\rm p}(t|{\bf
x},t')-f^{-1}_{\rm p}{\cal L}_{\rm e}(f_{\rm p})({\bf x},t')
B'_{\rm p}(t|{\bf x},t')
\end{eqnarray}
with final condition $B'_{\rm p}(t|{\bf x},t)=B({\bf x})$. The
proof is obvious if one employs the evolution equation in the
first line and the adjoint characteristic of ${\cal L}_{\rm e}$
and ${\cal L}^+_{\rm e}$. Actually these operators could be
arbitrary. Regarding the second term in the second line in
Eq.~(\ref{mutevoleqofdynamicobs}) as a small perturbation and
employing previous either approach, we will obtain
Eq.~(\ref{linearexpansionFDR}) again. This discussion also leads
into another interesting result. In physics the identification
between the perturbed and unperturbed systems is some arbitrary.
One may think of the unperturbed system as an oppositely perturbed
consequence of the perturbed system, e.g. applying mechanic forces
with opposite directions starting from time $0$. This point is
very clear in Eq.~(\ref{linearapproximation}), where $B_1$ of
course can be moved to the left hand side. Correspondingly, we
have an equation about $B'_{\rm o}(t|{\bf x},t')$ that is a
variation of the dynamic variable $B_{\rm o}(t|{\bf x},t')$ given
by
\begin{eqnarray}
\label{mutevoleqofdynamicobsO}
\partial_{t'} B'_{\rm o}(t|{\bf x},t')&=&-{\cal L}^{+}_{\rm o}({\bf x},t')B'_{\rm o}(t|{\bf x},t')
-f^{-1}_{\rm o}({\bf x},t')\left[f_{\rm o}{\cal L}^+_{\rm e}
-{\cal L}_{\rm e}(f_{\rm o})\right]({\bf x},t') B'_{\rm o}(t|{\bf
x},t')\nonumber\\
&=&-{\cal L}^{+}_{\rm p}({\bf x},t')B'_{\rm o}(t|{\bf
x},t')+f^{-1}_{\rm o}{\cal L}_{\rm e}(f_{\rm o})({\bf x},t')
B'_{\rm o}(t|{\bf x},t')
\end{eqnarray}
with final condition $B'_{\rm o}(t|{\bf x},t)=B({\bf x})$. One can
obtain it as well from Eq.~(\ref{mutevoleqofdynamicobs}) by simply
exchanging the subscripts ``p" and ``o" and changing the symbols
before ${\cal L}^+_{\rm e}$ and ${\cal L}_{\rm e}$ into minus.
Repeating previous evaluation, one obtains
Eq.~(\ref{linearexpansionFDR}) again. A more intriguing fact
appears when we tried to prove the Chapman-Kolmogorov
equation~(\ref{ChapmanKolmogorovFDcase}) for the function $B'_{\rm
p}$ using the evolution equation in the second line of
Eq.~(\ref{mutevoleqofdynamicobs}): vanishing of the derivative
with respect to time $t'$ on the left hand side of
Eq.~(\ref{ChapmanKolmogorovFDcase}) requires
\begin{eqnarray}
{\cal L}_{\rm e} (f_{\rm p})({\bf x},t')=[\partial_{t'}-{\cal
L}_{\rm o}]f_{\rm p}({\bf x},t').
\end{eqnarray}
It is obvious if we employ the forward equation for the
distribution $f_{\rm p}$. But this point reminds us a general
result: for an \emph{arbitrary} probability distribution $f({\bf
x},t)$ we can construct a function $B(t|{\bf x},t')$ satisfying a
perturbed backward equation
\begin{eqnarray}
\label{evoleqofdynamicobsA}
\partial_{t'} B(t|{\bf x},t')=
-{\cal L}^{+} ({\bf x},t')B(t|{\bf x},t') -f^{-1}({\bf
x},t')\left[\partial_{t'}f-{\cal L} (f) \right]({\bf x},t')
B(t|{\bf x},t')
\end{eqnarray}
with final condition $B'(t|{\bf x},t)=B({\bf x})$, and this
function satisfies
\begin{eqnarray}
\label{generlizedChapmanKolmogorovEq} \partial_{t'}[\int d{\bf
x}B(t|{\bf x},t')f({\bf x},t')]=0.
\end{eqnarray}
We call Eq.~(\ref{generlizedChapmanKolmogorovEq}) generalized
Chapman-Kolmogorov equation because the functions therein may be
beyond those in the standard one~(\ref{ChapmanKolmogorovFDcase}).
Eq.~(\ref{evoleqofdynamicobsA}) is not yet the most general; one
can still add new terms as those in
Eq.~(\ref{mutevoleqofdynamicobs}) to obtain other equations, which
will be seen shortly.

So far we employed the perturbation technique to solve the
backward equations~(\ref{evoleqofdynamicobs}) and reobtained the
linear response theory. Equations~(\ref{mutevoleqofdynamicobs})
and~(\ref{mutevoleqofdynamicobsO}) seem unnecessary because they
are not beyond the original one from the point of view of
perturbation. However, the Feynman-Kac formula~(\ref{FKformula})
and generalized Chapman-Kolmogorov
equation~(\ref{generlizedChapmanKolmogorovEq}) provide us two
nonperturbative relations:
\begin{eqnarray}
\label{nonperturbationFDRFK} \langle B\rangle_{\rm p}(t)&=&
 \langle\exp[\int_0^tf^{-1}_{\rm p}{\cal L}_{\rm e}(f_{\rm p})(\tau,{\bf
x}(\tau))d\tau]B({\bf x}(t)) \rangle_{\rm o}d\tau,\\
\label{nonperturbationFDRFKO} \langle B\rangle_{\rm o}(t)&=&
 \langle\exp[-\int_0^tf^{-1}_{\rm o}{\cal L}_{\rm e}(f_{\rm o})(\tau,{\bf
x}(\tau))d\tau]B({\bf x}(t)) \rangle_{\rm p}d\tau.
\end{eqnarray}
There is an analogous relation for Eq.~(\ref{evoleqofdynamicobsA})
as well. We must emphasize that these relations are always correct
formally and do not matter with the type of the perturbations.
Particularly, Eqs.~(\ref{nonperturbationFDRFK})
and~(\ref{nonperturbationFDRFKO}) reduce to the linear response
formula~(\ref{linearexpansionFDR}) when expanding their
exponentials to the first order. In addition to the Feynman-Kac
formula, we also notice that the Girsanov
formula~(\ref{CMGformula}) presents an alternative nonperturbative
relation for the perturbation problem,
\begin{eqnarray}
\label{nonperturbationFDRCMG} \langle B\rangle_{\rm p}(t)= \langle
B({\bf x}(t))\rangle_{\rm p}=\langle e^{-\int_0^t {\cal R}[{\bf
a}]({\bf x}(\tau))d\tau}B({\bf x}(t)) \rangle_{\rm
 o},
\end{eqnarray}
where ${\bf a}=h{\bf M}\nabla{V}$ for the mechanical perturbation
in the previous overdamped Brownian motion. At first sight, one
may think of that Eq.~(\ref{nonperturbationFDRCMG}) is superior to
Eq.~(\ref{nonperturbationFDRFK}) in that the latter does not need
the unknown perturbed distribution function $f_{\rm p}$. However,
Eq.~(\ref{nonperturbationFDRCMG}) is based on the validity of the
Girsanov formula. We have mentioned that this formula is not
always true, e.g., the general perturbation in the underdamped
Brownian motion; see a simple discussion in
Appendix~\ref{extendedCMGformula}. In contrast,
Eq.~(\ref{nonperturbationFDRFK}) is robust.

Both Eqs.~(\ref{nonperturbationFDRFK})
and~(\ref{nonperturbationFDRCMG}) have to face a challenge whether
they are really useful, which relies on whether they provide us
new evaluation approaches or physical understanding about
stochastic processes. It should be better to put this question
into a more general situation, namely, whether
Eq.~(\ref{evoleqofdynamicobsA}) is useful or not. This is natural
because Eqs.~(\ref{mutevoleqofdynamicobs})
and~(\ref{mutevoleqofdynamicobsO}) are its specific cases. We have
mentioned that even Eq.~(\ref{evoleqofdynamicobsA}) has a more
general variant,
\begin{eqnarray}
\label{evoleqofdynamicobsG}
\partial_{t'} B(t|{\bf x},t')&=&-{\cal L}^{+}({\bf x},t') B(t|{\bf x},t')
-f^{-1}({\bf x},t')\left[ \partial_{t'} f- {\cal L}(f) \right]({\bf
x},t') B(t|{\bf x},t') \\&&+f^{-1}({\bf x},t')\left[{\cal L}_{\rm
a}(g)-g{\cal L}_{\rm a}^{+}\right]({\bf x},t') B(t|{\bf
x},t'),\nonumber
\end{eqnarray}
with final condition $B(t|{\bf x},t)=B({\bf x})$, where $g({\bf
x},t')$ is arbitrary smooth positive functions, the arbitrary
operators ${\cal L}_a$ and ${\cal L}_a^{+}$ are adjoint each
other. One may check that, under this case the generalized
Chapman-Kolmogorov equation~(\ref{generlizedChapmanKolmogorovEq})
is still true. Intriguingly, the two perturbed components in
Eq.~(\ref{evoleqofdynamicobsG}) have very distinct meanings for
the generalized Chapman-Kolmogorov equation: the first in the
first line is indispensable while the second in the second line is
not. This point should be reflected in the physical explanations
of the above equation. Rather than investigating very general
${\cal L}_a$, in this work we are interested in the simplest but
nontrivial case: ${\cal L}_{\rm a}$ and $g({\bf x},t)$ are chosen
such that Eq.~(\ref{evoleqofdynamicobsG}) is
\begin{eqnarray}
\label{evoleqofdynamicobsGS} \cdots+2f^{-1}({\bf
x},t')\left[(\partial_{x_i} S_i)({\bf x},t')+S_i({\bf
x},t')\partial_{x_i}\right] B(t|{\bf x},t'),
\end{eqnarray}
where ``$\cdots$" represents the first line of
Eq.~(\ref{evoleqofdynamicobsG}), and $N$-dimension vector ${\bf
S}$$=$$\{S_i\}$ satisfies natural boundary condition. On the basis
of the generalized Chapman-Kolmogorov equation, Feynman-Kac and
limited Girsanov formulas (Appendix~\ref{extendedCMGformula}), for
a certain vector ${\bf S}$ whose last $(N$$-$$M)$ components
vanish, we obtain an identity
\begin{eqnarray}
\langle e^{-\int_0^t{\cal J}[f,{\bf S}]({\bf x}(\tau),\tau)d\tau
}B[{\bf x}(t)]\rangle=\langle B\rangle(t),\label{GIFT}
\end{eqnarray}
where the integrand is
\begin{eqnarray}
\label{functional} {\cal J}[f,{\bf S}]&=& f^{-1}\left[\left({\cal
L}-\partial_\tau\right)f+2\partial_{x_i}S_i\right] +{{\cal R}}[-2f^{-1}{\bf S}]\nonumber\\
&=&f^{-1}\left[\left({\cal
L}-\partial_\tau\right)f+2\partial_{x_i}S_i +2f^{-1} S_i({\bf
B}^{-1})_{il} S_l\right] +2f^{-1}S_i({\bf
B}^{-1})_{il}\left(v_l-A_l \right),
\end{eqnarray}
the inverse of ${\bf B}$ is formally defined by
\begin{eqnarray}
\label{FormallyinversedB}
 \bf{B}^{-1}=\left[
  \begin{array}{cc}
    \bf{D}^{-1} &0\\
    0 & 0 \\
  \end{array}
\right],
\end{eqnarray}
the mean on the left hand side is over the trajectories starting
from initial distribution function $f({\bf x},0)$ and determined
by the stochastic process~(\ref{SDE}), and the mean on the right
hand side denotes the average over distribution $f({\bf x},t)$. We
call Eq~(\ref{GIFT}) generalized integral fluctuation theorem,
which is obviously more general than previous version that was
limited to the Smoluchowski type diffusions~\cite{LiuF1}. Noting
time $0$ in the GIFT may be replaced by any time $t'$ ($<$$t$) and
correspondingly the average on the left hand side is over  $f({\bf
x},t')$.

\subsection{GIFT and time reversal}\label{GIFTtimereversal}
As mentioned at the very beginning, the backward equation has a
natural connection with time reversal. A naive understanding about
it may define a reversed time $s$$=$$t$$-$$t'$ ($0$$\leq$
$t'$$\leq t$) and convert the backward equation into initial value
problem. This would be useful when applying ordinary numerical
approaches to the unusual final value problem. However, the
situation is more delicate about time reversal of
Eq.~(\ref{evoleqofdynamicobsGS}). Multiplying both sides of the
equation by $f({\bf x},t')$ and performing a simple
reorganization, we obtain
\begin{eqnarray}
\partial_{t'}[B(t|{\bf x},t')f({\bf x},t')]&=&
-f({\bf x},t'){\cal L}^{+}B(t|{\bf x},t')+{\cal L}(f)({\bf
x},t')B(t|{\bf x},t') + 2 \partial_{x_i}[ S_i({\bf x},t') B(t|{\bf
x},t')].
\end{eqnarray}
Compared with Eq.~(\ref{identity3}), we see that, if choosing
$S_i$ to be the probability current $J_i(f)$ the right hand side
becomes $-{\cal L}[B(t|{\bf x},t')f({\bf x},t')]$. Using the new
time parameter $s$ rather than $t'$, we then obtain a time
reversed Fokker-Planck equation for function $B(t|{\bf
x},t')f({\bf x},t')$ and the Fokker-Planck operator is simple
${\cal L}({\bf x},t-s)$. This argument was further generalized to
the case with even and odd variables ${\bf x}$ under time
reversal~\cite{LiuF1}. Because the stochastic process~(\ref{SDE})
here is more general than previous one, and time reversal is very
important in following discussions, e.g. the derivation of
transient DFT, we briefly recall some definitions and main
results.

Coordinates $x_i$ of stochastic system may be even or odd,
according to their rules under time reversal: if $x_i$$\to$$ +x_i$
is even and $x_i$$\to$$-x_i$ is odd,  e.g., momentum in
Eq.~(\ref{underampedBM}); in abbreviation $x_i$$\to$$\tilde{x}_i$$
=$$\varepsilon_i x_i$ and $\varepsilon_i$$=\pm1$. The drift vector
splits into ``irreversible" and ``reversible" parts, ${\bf A}={\bf
A}^{\rm irr}+{\bf A}^{\rm rev}$. Under a time reversal, we assume
these vectors are transformed into $\tilde{{\bf A}}=\tilde{{\bf
A}}^{\rm irr}+{\tilde{\bf A}}^{\rm rev}$, where
\begin{eqnarray}
\label{reverseddrifts} \tilde{A}_i^{\rm irr}({\bf x},
t')&=&\varepsilon_iA_i^{\rm
irr}({\bf\tilde{x}}, s),  \\
\tilde{A}_i^{\rm rev}({\bf x}, t')&=&-\varepsilon_iA_i^{\rm
rev}({\bf \tilde{x}}, s).
\end{eqnarray}
Such a splitting may be arbitrary or a prior known. Additionally,
the transformation of the diffusion matrix is also given by
\begin{eqnarray}
\label{reverseddiffcoeff} \tilde{B}_{il}({\bf
x},t')&=&\varepsilon_i\varepsilon_l B_{il}({\bf \tilde{x}}, s).
\end{eqnarray}
No summation over repeated indices here. These transformations are
actually an inhomogeneous extension of homogeneous diffusion
case~\cite{Risken,Gardiner}. Considering a time reversed forward
Fokker-Planck equation with above new defined drift vector and
diffusion matrix,
\begin{eqnarray}
\label{ReversedFPoperator}
\partial_s p({\bf \tilde{x}},s)={\cal L}_{\rm R}({\bf \tilde{x}},s)p({\bf \tilde{x}},s)
=[-\partial_{\tilde{x}_i} {\tilde A}_{i}({\bf
\tilde{x}},s)+\frac{1}{2}\partial_{\tilde{x}_i}\partial_{\tilde{x}_j}{\tilde
B}_{ij}({\bf \tilde{x}},s)]p({\bf \tilde{x}},s).
\end{eqnarray}
Substituting a decomposition
\begin{eqnarray}
p({\bf \tilde{x}},s)=[\int b(t|{\bf x}',t)f({\bf x}',t)d{\bf
x}']^{-1}b(t|{\bf x},t')f({\bf x},t'), \label{decompositionTR}
\end{eqnarray}
where $b(t|{\bf x},t)$$=$$B({\bf x})$, $f({\bf x},t')$ is an
arbitrary normalized positive function, and the prefactor ensures
$p({\bf \tilde{x}},0)$ to be normalized, and a performing simple
evaluation, we can rewrite Eq.~(\ref{ReversedFPoperator}) as
\begin{eqnarray}
\label{evoleqofdynamicobsGSCurrent}
\partial_{t'} b(t|{\bf x},t')&=&-{\cal L}^{+}({\bf x},t') b(t|{\bf x},t')
-f^{-1}({\bf x},t')\left[ \partial_{t'} f- {\cal L}(f)
\right]({\bf x},t') b(t|{\bf x},t') \\&&+2f^{-1}({\bf
x},t')\left[(\partial_{x_i} S^{\rm irr}_i(f)) +S^{\rm
irr}_i(f)\partial_{x_i}\right] b(t|{\bf x},t'),\nonumber
\end{eqnarray}
where we define an irreversible probability current on the
function $f$
\begin{eqnarray}
\label{irrcurrent} S_i^{\rm irr}(f)= A^{\rm irr}_i({\bf
x},t')f({\bf x},t')-\frac{1}{2}
\partial_{x_l}(B_{il}f)({\bf x},t').
\end{eqnarray}
Hence, if vector ${\bf S}$ in Eq.~(\ref{evoleqofdynamicobsGS})
equals the irreversible current, the time reversal explanation of
the equation is just Eq.~(\ref{ReversedFPoperator}). Moreover,
this explanation is still valid even in case of general ${\bf S}$.
One may easily see it by constructing a specific splitting
\begin{eqnarray}
\label{splitting} A^{\rm irr}_i({\bf x},t|f,{\bf S})&=&f^{-1}({\bf
x},t)
[S_i({\bf x},t)+\frac{1}{2}\partial_{x_l}(B_{il}f)({\bf x},t) ],\\
A^{\rm rev}_i({\bf x},t|f,{\bf S})&=&A_i({\bf x},t)-A_i^{\rm
irr}({\bf x},t|f,{\bf S}).
\end{eqnarray}
Obviously, ${\bf S}$ is just the irreversible probability current
defined by the above irreversible drift on function $f$, which we
denote ${\bf S}^{\rm irr}(f|{\bf S},f)$ in the following. We must
emphasize that such a splitting might be not real in physics. The
relationship between Eqs.~(\ref{evoleqofdynamicobsGS})
and~(\ref{ReversedFPoperator}) presents an alternative
understanding of the generalized Chapman-Kolmogorov
equation~(\ref{generlizedChapmanKolmogorovEq}): the spatial
integral of its left hand side is proportional to the total
probability of $p(\tilde{\bf x},s)$ that is time-invariable
according to the forward equation~(\ref{ReversedFPoperator}). It
is worth emphasizing that the above conclusions do not matter with
the characteristics of the diffusion matrix (degenerate or
nondegenerate). We believe that we should not be the first to
obtain Eq.~(\ref{evoleqofdynamicobsGSCurrent}). This equation
might be derived earlier in finding the conditions on the
diffusion matrix and drift vector for a time-reversible
homogeneous Fokker-Planck equation ($\tilde{{\bf A}}^{\rm
irr}$$=$${\bf A}^{\rm irr}$, ${\tilde{\bf A}}^{\rm rev}$$=$${\bf
A}^{\rm rev}$, and $\tilde{\bf B}$$=$${\bf B}$) to have stationary
equilibrium solution $f^{\rm eq}({\bf x})$ that satisfies the
detailed balance principle~\cite{Gardiner,Risken}. We see these
conditions are identical to the requirement that $f$$=$$f^{\rm
eq}({\bf x})$ and the other terms except for ${\cal L}^+$ on the
right hand side of Eq.~(\ref{evoleqofdynamicobsGSCurrent}) vanish,
respectively.

\subsection{GIFT and integral transient fluctuation theorems}
Although the GIFT~(\ref{GIFT}) is always correct in mathematics,
their physical meaning and applications in practice are not very
obvious given very general $f$ and ${\bf S}$. These problems might
be answered better by choosing familiar functions with explicit
physical meaning, e.g., probability distribution function and
irreversible probability current of stochastic system, or choosing
very simple expressions. We have briefly reported
that~\cite{LiuF1}, under some specific choices the GIFT reduced to
existing several
IFTs~\cite{JarzynskiPRE97,JarzynskiPRL97,Maes,SeifertPRL05,HatanoSasa}.
Here we present detailed evaluations, and particularly we add the
results about the Kramers diffusion and the new
IFTs~(\ref{nonperturbationFDRFK})
and~(\ref{nonperturbationFDRFKO}). One will see the GIFT actually
provides a simple and clear way to classify these IFTs.

\subsubsection{${\bf S}={\bf S}^{\textrm irr}(f)$ with natural splitting}
\label{priorsplitting} If we prior know a splitting of the drift
vector, this may be the most natural consideration. In the
derivation of Eq.~(\ref{evoleqofdynamicobsGSCurrent}) from the
time reversed Fokker-Planck equation~(\ref{ReversedFPoperator}),
function $f({\bf x},t')$ in the
decomposition~(\ref{decompositionTR}) is almost arbitrary. One may
specify a decomposition $p({\bf y},s)$$\propto$$ 1\times b_{(\rm
1)}(t|{\bf x},t')$ and the new function $b_{(\rm 1)}(t|{\bf
x},t')$ still satisfies Eq.~(\ref{evoleqofdynamicobsGSCurrent})
except for $f=1$ therein. Because of the same $p({\bf y},s)$,
these two decomposition has a simple connection,
\begin{eqnarray}
\label{entropybalancefunction} b(t|{\bf x},t')=\frac{b_{(\rm
1)}(t|{\bf x},t')}{f({\bf x},t')}\frac{\int b(t|{\bf x}',t) f({\bf
x}',t)d{\bf x}' }{\int b_{(\rm 1)}(t|{\bf x}',t)d{\bf x}' }.
\end{eqnarray}
This result immediately results into a relationship between the
functionals~(\ref{GIFT}) of the path integral representations of
$b(t|{\bf x},0)$ and $b_{\rm 1}(t|{\bf x},0)$:
\begin{eqnarray}
\label{entropybalancefunctional} \int_0^t{\cal J}[f,{\bf S}^{\rm
irr}(f)]({\bf x}(\tau),\tau)d\tau=-\ln \frac{f({\bf
x}(t),t)}{f({\bf x}(0),0)}+\int_0^t{\cal J}_{(\rm 1)}({\bf
x}(\tau),\tau)d\tau
\end{eqnarray}
where ${\cal J}_{(\rm 1)}$$=$${\cal J}[1,{\bf S}^{\rm irr}(1)]$,
and the term $\ln f({\bf x}(t),t)$ is from the final condition
$b_{(\rm 1)}(t|{\bf x},0)$. Given a prior known splitting ${\bf
A}={\bf A}^{\rm irr}+{\bf A}^{\rm rev}$ and performing a simple
evaluation, the new function has an expression
\begin{eqnarray}
\label{functional1} {\cal J}_{(\rm
1)}&=&\partial_{x_i}\hat{A}_i^{\rm irr}-\partial_{x_i} A_i^{\rm
rev}+ 2 \hat{A}_i^{\rm irr}({\bf B}^{-1})_{il}\hat{A}_l^{\rm irr}
+2\hat{A}_i^{\rm irr}({\bf B}^{-1})_{il}(v_l-A_l)\nonumber\\
&=&2 \hat{A}_i^{\rm irr}({\bf B}^{-1})_{il}(v_l-A_l^{\rm
rev})-\partial_{x_i} A_i^{\rm rev} \hspace{0.5cm}({\rm S}),
\end{eqnarray}
where $\hat{A}_i^{\rm irr}=A_i^{\rm irr}-\partial_{x_l}B_{il}/2$,
and letter ``S" in the second line denotes that time integral of
this equation is Stratonovich integral~\cite{comment1}. Compared
with the original one, function $b_1(t|{\bf x},t')$ is distinctive
because its functional is completely determined by intrinsic
characteristics of the system and environment, including the drift
vector and diffusion matrix. Moreover, the above functional
identity~(\ref{entropybalancefunctional}) definitely states that,
for any pair of functions having the same expressions at times $0$
and $t$, their GIFTs under this consideration are completely
identical. An analogous expression was obtained earlier in
Ref.~\cite{Chetrite08CMP} [Eq.~(7.5) therein] by using an abstract
time reversal argument. We may emphasize that
Eq.~(\ref{functional1}) is more general than the previous one,
because it also accounts for Kramers diffusion, which is seen
shortly.

Equation~(\ref{functional1}) has simpler expressions for the two
physical models in Sec.~\ref{linearresponsetheory}. For the
overdamped Brownian motion~(\ref{multidimensionaloverdampedBM})
with even variables only ($\varepsilon_i$$=$$+$), a conventional
splitting is
\begin{eqnarray}
\label{reversedprotocol} {\bf A}^{\rm irr}={\bf A}({\bf
x},t),\hspace{0.3cm}{\bf A}^{\rm rev}({\bf x},t)=0.
\end{eqnarray}
Then we have ${\bf S}^{\rm irr}(f)$$=$${\bf J}(f)$. The time
reversal of this splitting was called reversed
protocol~\cite{Chernyak}. Correspondingly, if the mobility matrix
and the environment temperature are constant, ${\cal J}_{(\rm 1)}$
is simply
\begin{eqnarray}
\beta (-\partial_{x_i}U+{\bf F}_i)v_i \label{functionalODampedBM}
\end{eqnarray}
Another example is the underdamped Brownian
motion~(\ref{underampedBM}). Different from the overdamped case,
this model has even spatial coordinate and odd momentum
coordinate. For a simple Hamiltonian ${{\cal H}_0}=
p^2/2m+U(x,t)$, we have a canonical splitting
\begin{eqnarray}
\label{Kramerssplitting}
 {\bf A}^{\rm irr}({\bf r},t)=-\Gamma\cdot{\bf
P},\hspace{0.3cm} {\bf A}^{\rm rev}({\bf r},t)=\Pi\cdot\nabla
{{\cal H}_0}+{\bf F}.
\end{eqnarray}
Then $p$-component of the irreversible current on function $f$ is
\begin{eqnarray}
\label{Kramersirrcurrent} S^{\rm irr}_p(f)&=&-\gamma_0p
f(p,x)-\partial_p[\beta^{-1}m\gamma_0f(p,x)],
\end{eqnarray}
and $x$-component $S^{\rm irr}_x(f)$ vanishes. Therefore, the
condition for the GIFT~(\ref{GIFT}) with degenerate diffusion
matrix is satisfied. Under an assumption of constant friction
coefficient and environment temperature, ${\cal J}_{(\rm 1)}$ is
simplified into
\begin{eqnarray}
-\beta\frac{d}{d\tau}(\frac{mv^2}{2})+\beta(-\partial_x U+F)v.
\label{functionalUDampedBM}
\end{eqnarray}
We see the overdamped result~(\ref{functionalODampedBM}) can be
obtained by letting $m=0$ in the above equation. If the
temperature is a function of spatial coordinate, one may easily
check that the time integral of ${\cal J}_{(\rm 1)}$ is Eq.~(6.12)
in Ref.~\cite{Lebowitz} that was called entropy flow from the
system to environment along a trajectory.

The physical meaning of
functional~(\ref{entropybalancefunctional}) has been well
understood~\cite{Maes,SeifertPRL05,Chetrite08CMP}: If the function
$f$ is the probability distribution function $\rho({\bf x},t)$ of
the stochastic system, the first and second terms are the Gibbs
entropy production of the system and the entropy production in
environment along a stochastic trajectory between times $0$ and
$t$, respectively. Hence the GIFT~(\ref{GIFT}) under this
consideration is the IFT of the overall entropy production given a
specific splitting. This theorem also presents that, for a
diffusion process the mean overall entropy production of
stochastic system is always nonnegative (the second law of
thermodynamics). This point may be seen by directly using Jensen
inequality to the GIFT with $B$$=$$1$ or evaluating the mean
instantaneous rate of overall entropy production, the latter of
which is
\begin{eqnarray}
\label{intantaneousentropyproduction} \langle{\cal J}[\rho,{\bf
S}^{\rm irr}(\rho)]\rangle=2\int d{\bf x} \rho^{-1} S^{\rm
irr}_i(\rho)({\bf B}^{-1})_{il} S^{\rm irr}_l(\rho)\geq0.
\end{eqnarray}
Noticing that the other terms in Eq.~(\ref{functional}) all vanish
after ensemble average (the last term due to the definition of Ito
integral~\cite{Gardiner}). Noting
Eq.~(\ref{intantaneousentropyproduction}) also holds for any
vector ${\bf S}$ with natural boundary condition.

\subsubsection{Vanishing $\bf S$ with posterior splitting}
For an arbitrary vector ${\bf S}$, the above
results~(\ref{entropybalancefunction})-(\ref{functional1}) are
still correct except that they are about $B(t|{\bf x},t')$ and
$B_{(\rm 1)}(t|f,{\bf S},{\bf x},t')$ and their functionals, where
the decomposition $p({\bf y},s)$ $\propto $$1\times B_{(\rm
1)}(t|f,{\bf S},{\bf x},t')$. Significantly different from
previous case, both $B_{(\rm 1)}$ and its ${\cal J}_{(\rm 1)}$
depend on $f$ and ${\bf S}$ through the
splitting~(\ref{splitting}). Because such a splitting is defined
under these given functions, we roughly call it posterior. Rather
than discussing a general vector, we focus on the simplest case
${\bf S}=0$. Correspondingly, the splitting is
\begin{eqnarray}
\label{Svanishingsplitting} A^{\rm irr}_i({\bf x},t|f)=\frac{1}{2f
({\bf x},t)}\partial_{x_l}(B_{il}f)({\bf
x},t),\hspace{0.3cm}A^{\rm rev}_i({\bf x},t|f)=A_i({\bf
x},t)-A^{\rm irr}({\bf x},t|f).
\end{eqnarray}
Substituting them into Eq.~(\ref{functional1}), we obtain
\begin{eqnarray}
{\cal J}_{(\rm 1)}[f]=f^{-1}({\cal L}+v_i\partial_{x_i})f
\hspace{0.3cm}({\rm S}).
\end{eqnarray}
The same result can be achieved simply by employing the relation
$d/d{\tau}$$=$$\partial_{\tau}$$+$$v_i\partial_{x_i}$ and
\begin{eqnarray}
\label{entropybalancefunctionalS0} {\cal J}[f,0]= f^{-1}({\cal
L}-\partial_\tau) f.
\end{eqnarray}

Equation~(\ref{Svanishingsplitting}) shows a posterior splitting
is usually $f$-dependence. But there is an intriguing exception if
the drift vector and diffusion matrix of a stochastic system
satisfy the detail balance conditions when time parameters in them
are fixed. Such a system has a transient equilibrium solution
\begin{eqnarray}
\label{transientdetailcondition} {\cal L}({\bf x},t)f^{\rm
eq}({\bf x},t)=0, {\hspace{0.3cm}}{\bf S}^{\rm irr}(f^{\rm eq})=0.
\end{eqnarray}
and this solution has a simple Boltzmann distribution. For
instance, in the models~(\ref{multidimensionaloverdampedBM})
and~(\ref{underampedBM}) with constant mobility matrix and
friction coefficient, if nonconservative forces there vanish, such
solutions indeed exist and $f^{\rm eq}$ $\propto$ $\exp[-\beta U]$
and $\propto$ $\exp[-\beta (p^2/2m+U)]$, respectively. Hence, if
we choose $f=f^{\rm eq}({\bf x},t)$, the
splitting~(\ref{Svanishingsplitting}) is no longer $f$-dependent
and Eq.~(\ref{entropybalancefunctionalS0}) becomes
\begin{eqnarray}
\label{entropybalancefunctionalJE} {\cal J}_{\rm JE}[f^{\rm eq},0]
=-\partial_\tau \ln f^{\rm eq}.
\end{eqnarray}
The time integral of the above equation was called the dissipated
work. One easily sees that, under this case the GIFT~(\ref{GIFT})
with $B$$=$$1$ and $\delta({\bf x}-{\bf z})$ are the celebrated
Jarzynski equality~\cite{JarzynskiPRE97,JarzynskiPRL97} and the
key Eq.~(4) in the Hummer and Szabao's work~\cite{Hummer01},
respectively. Although the splitting here is the same with the
natural splitting we discussed previously, we must point out that,
in the Jarzynski equality, stochastic trajectories start from an
initial equilibrium distribution. In contrast, the IFT of the
overall entropy production is valid for any initial distribution
besides equilibrium state and even in the presence of
nonconservative forces.

A famous example of virtually $f$-dependent splitting in the
literature is for the stochastic system having transient
nonequilibrium steady-state~\cite{HatanoSasa},
\begin{eqnarray}
\label{transientsteadycondition} {\cal L}({\bf x},t)f^{\rm
ss}({\bf x},t)=0, \hspace{0.3cm} {\bf J} (f^{\rm ss})\neq 0.
\end{eqnarray}
e.g., nonconservative forces nonzero in the
models~(\ref{multidimensionaloverdampedBM})
and~(\ref{underampedBM}). The time reversal corresponding the
splitting of Eq.~(\ref{Svanishingsplitting}) with $f$$=$$f^{\rm
ss}$ was also called current reversal~\cite{Chetrite08CMP}.
Equation~(\ref{entropybalancefunctionalS0}) under this case
becomes
\begin{eqnarray}
\label{entropybalancefunctionalHS} {\cal J}_{\rm HS}[f^{\rm
ss},0]=-\partial_\tau \ln f^{\rm ss}.
\end{eqnarray}
We see it is almost the same with
Eq.~(\ref{entropybalancefunctionalJE}) thought their splitting or
time reversals are completely different. The time integral of the
above equation was called the excess heat or entropy production
and the GIFT with $B$$=$$1$ is the Hatano-Sasa
equality~\cite{HatanoSasa}. Noting stochastic trajectories of this
theorem start from a nonequilibrium steady-state.

In addition to the above two well-known IFTs,
Eq.~(\ref{entropybalancefunctionalS0}) also reveals several
simpler IFTs with vanishing ${\bf S}$. The most obvious case is to
choose $f$$=$$\rho$ the distribution function of the stochastic
system itself and ${\cal J}[\rho,0]$$=$$0$ simply.
Correspondingly, Eq.~(\ref{evoleqofdynamicobsGS}) reduces to the
standard Kolmogorov backward equation~(\ref{Kbackoperator}) and
now the GIFT~(\ref{GIFT}) is trivially the path integral
representation of the standard Chapman-Kolmogorov
equation~(\ref{ChapmanKolmogorov}); also see
Eqs~(\ref{defdynamicobserval}),~(\ref{mean&dynamicobserval})
and~(\ref{ChapmanKolmogorovFDcase}). The splitting or time
reversal~(\ref{Svanishingsplitting}) in this case was called
complete reversal~\cite{Chetrite08CMP}. The other IFTs are
relevant to the perturbation problem in
Sec.~(\ref{linearresponsetheory}). We choose the stochastic
systems to be the unperturbed one ${\cal L}$$=$${\cal L}_{\rm o}$
and $f$$=$$f_{\rm p}$ or the perturbed one ${\cal L}$$=$${\cal
L}_{\rm p}$ and $f$$=$$f_{\rm o}$ as discussed previously,
Eq.~(\ref{entropybalancefunctionalS0}) then becomes
\begin{eqnarray}
\label{entropybalancefunctionalS0FDR} {\cal J}[f_{\rm
p},0]&=&f_{\rm p}^{-1}({\cal L}_{\rm o}-\partial_\tau f_{\rm
p})=-f^{-1}_{\rm p}{\cal L}_{\rm e}(f_{\rm p}),\\
\label{entropybalancefunctionalS0FDRO} {\cal J}[f_{\rm
o},0]&=&f_{\rm o}^{-1}({\cal L}_{\rm p}-\partial_\tau f_{\rm
o})=f^{-1}_{\rm o}{\cal L}_{\rm e}(f_{\rm o }),
\end{eqnarray}
respectively. We immediately see that the corresponding GIFTs are
Eqs.~(\ref{nonperturbationFDRFK})
and~(\ref{nonperturbationFDRFKO}), respectively. Although these
identities look very similar, their time reversals definition are
significantly different. Let us consider a simple situation that
the unperturbed system is in equilibrium $f^{\rm eq}_{\rm o}({\bf
x})$ and the perturbation ${\bf A}_{\rm e}({\bf x},t)$ is imposed
on the drift vector ${\bf A}_{\rm o}({\bf x})$$=$${\bf A}^{\rm
rev}_{\rm o}({\bf x})+{\bf A}_{\rm o}^{\rm irr}({\bf x})$ as
usual. Obviously, for the case ${\cal L}$$=$${\cal L}_{\rm o}$ and
$f$$=$$f_{\rm p}$, the posterior
splitting~(\ref{Svanishingsplitting}) is $f_{\rm p}$-dependence.
We usually do not know their concrete expressions due to the
unknown $f_p$. On the contrary, for the case ${\cal L}$$=$${\cal
L}_{\rm p}$ and $f$$=$$f_{\rm o}$, because $f^{\rm eq}_{\rm
o}({\bf x})$ satisfies the detailed balance condition,
Eq.~(\ref{Svanishingsplitting}) is simply
\begin{eqnarray}
\label{SvanishingsplittingFD}{\bf A}^{\rm irr}({\bf x},t|f_{\rm
o})={\bf A}^{\rm irr}_{\rm o}({\bf x}),\hspace{0.3cm}{\bf A}^{\rm
rev}({\bf x},t|f_{\rm o})={\bf A}^{\rm rev}_{\rm o}({\bf x})+{\bf
A}_{\rm e}({\bf x},t).
\end{eqnarray}
This is a new example with vanishing ${\bf S}$ and $f$-independent
time reversal particularly. Whatever the perturbation is
reversible or irreversible in physics, it is always classified
into the reversible drift in the time reversed system ${\cal
L}_{\rm R}$. This point is interesting for physical model with
vanishing ${\bf A}^{\rm rev}_{\rm o}$, e.g., the overdamped
Brownian motion~(\ref{multidimensionaloverdampedBM}) with
vanishing nonconservative force.

Different from Eq.~(\ref{intantaneousentropyproduction}), because
function $f$ is usually not identical to system's real
distribution function $\rho({\bf x},t)$, we cannot interpret the
ensemble average of Eq.~(\ref{entropybalancefunctionalS0}) as mean
instantaneous rate of overall entropy
production~(\ref{intantaneousentropyproduction}), though it is
always nonnegative (Jensen inequality). However, they are indeed
connected by the following relation,
\begin{eqnarray}
\label{relativeentropy} \int_0^t{\cal J}[f,{\bf S}^{\rm
irr}(f|f)=0]({\bf x}(\tau),\tau)d\tau=\ln \frac{\rho({\bf
x}(t),t)}{f({\bf x}(t),t)}+\int_0^t{\cal J}[\rho,{\bf S}^{\rm
irr}(\rho|f)]({\bf x}(\tau),\tau)d\tau,
\end{eqnarray}
where we have assumed $f$ and $\rho$ have the same distribution at
time $0$, the functional on the right hand side is for the new
function $D$ defined by a decomposition $p({\bf y},s)$$\propto$
$\rho({\bf x},t')D(t|{\bf x},t')$. We must emphasize that both the
time reversed Fokker-Planck equation for $p({\bf y},s)$ and the
irreversible probability current on the system's distribution
function $\rho$ here are constructed by the posterior
splitting~(\ref{Svanishingsplitting}).
Equation~(\ref{relativeentropy}) can be easily proved on the basis
of Eq.~(\ref{entropybalancefunctional}). Averaging both sides of
the above equation with respect to the distribution function
$\rho$, we see that the second term on the right hand side is the
mean overall entropy production during a fixed time $t$ given the
specific splitting~(\ref{Svanishingsplitting}), and the first term
is the relative entropy between the two distributions $\rho$ and
$f$ at time $t$, which is always nonnegative. Hence we call the
left hand side of Eq.~(\ref{relativeentropy}) overall relative
entropy production functional~\cite{Chetrite08CMP}. We may point
out that the above results are also suitable to the cases with
nonzero ${\bf S}$, e.g., see Eq.~(\ref{totalentropydecomp}) below.

\subsection{Girsanov equality}
\label{CMGequality} Recalling Eq.~(\ref{functional}), one may
notice that any ensemble average of the term $f^{-1} S_i({\bf
B}^{-1})_{il} S_l$ is always non-negative due to the semipositive
definite diffusion matrix ${\bf B}$. In fact, this observation has
alternative indirect explanation. Considering a perturbed forward
Fokker-Plank equation
\begin{eqnarray}
\partial_t\rho'={\cal L}'({\bf x},t)\rho'={\cal
L}({\bf x},t)\rho'+2\partial_{x_i}[f^{-1}({\bf x},t)S_i({\bf
x},t)\rho'({\bf x},t)].
\end{eqnarray}
Employing the limited Girsanov formula, we obtain an identity
\begin{eqnarray}
\label{CMGIFT}
 \langle e^{-\int_0^t {{\cal R}}[-2f^{-1}{\bf S}]({\bf x}(\tau),\tau)d\tau}B({\bf
 x}(t))\rangle=\langle B\rangle'(t),
\end{eqnarray}
and previous Eq.~(\ref{nonperturbationFDRCMG}) is its specific
case. We call the above equation with $B=1$ Girsanov equality.
Speck and Seifert first obtained such type of equality in a
specific case with ${\bf S}$$=$${\bf J}(f^{\rm ss})$ and
$f$=$f^{\rm ss}$ the transient steady-state defined in
Eq.~(\ref{transientsteadycondition})~\cite{Speck}. Jensen
inequality indicates the ensemble average of the functional of the
equality is nonnegative. It is worth emphasizing that
Eq.~(\ref{CMGIFT}) is related to the standard Chapman-Kolmogorov
equation~(\ref{differChapmanKolmogorov}) rather than the
generalized one~(\ref{generlizedChapmanKolmogorovEq}). This point
can be seen from the fact that the means of both sides are
respectively over $\rho'({\bf x},0)$ and $\rho'({\bf x},t)$ rather
than $f$ functions in the GIFT~(\ref{GIFT}).  This analysis also
reminds us an interesting relation given the vector ${\bf S}$
\emph{divergenceless}:
\begin{eqnarray}
\label{totalentropydecomp} \int_0^t{\cal J}[f,{\bf S}]({\bf
x}(\tau),\tau)d\tau &=&\ln \frac{\rho({\bf x}(t),t)}{f({\bf
x}(t),t)}+\int_0^t{\cal J}[\rho,{\bf S}^{\rm irr}(\rho|f,{\bf
S})]({\bf
x}(\tau),\tau)d\tau\\
&=&\int_0^t{\cal J}[f,0]({\bf x}(\tau),\tau)d\tau+\int_0^t {{\cal
R}}[-2f^{-1}{\bf S}]({\bf x}(\tau),\tau)d\tau.\nonumber
\end{eqnarray}
The first line is the version of Eq.~(\ref{relativeentropy}) for
nonzero ${\bf S}$, and the condition $\rho({\bf x},0)$$=$$f({\bf
x},0)$ was assumed. It is not difficult to find a nontrivial
divergenceless vector, e.g., ${\bf J}(f^{\rm ss})$ in the
overdamped Brownian motion~(\ref{multidimensionaloverdampedBM})
with nonzero time-dependent nonconservative force, which was also
the case investigated by Speck and Seifert~\cite{Speck}. Under
this consideration, choosing $f$ the transient steady-state and
further assuming the stochastic system to be in nonequilibrium
steady states $f^{\rm ss}({\bf x},t)$ at $t$, we find the first
line is just the overall entropy production functional of the
system, and the first term in the second line is the excess heat
or entropy production
functional~(\ref{entropybalancefunctionalHS}). Hence the last term
was called housekeeping heat functional to consist with
steady-state thermodynamics~\cite{Oono}.

\section{Transient detailed fluctuation theorem}
\label{GeneralizedCrooksrelation} The path integral representation
of the solution of Eq.~(\ref{evoleqofdynamicobsGS}) presents a
relationship between $B(t|{\bf x},t')$ with general final
condition and the one $B({\bf x}_2,t_2|{\bf x}_1,t_1)$ with
specific final condition $\delta({\bf x}_1-{\bf x}_2)$, which is
simply
\begin{eqnarray}
\label{ChapmanEqBackward} B(t|{\bf x}_1,t_1)&=&\int d{\bf
x}_2\langle\exp[-\int_{t_1}^{t_2}{\cal J}d\tau]\delta({\bf
x}(t_2)-{\bf x}_2)
\times\exp[-\int_{t_2}^t{\cal J} d\tau]B({\bf x}(t))\rangle\nonumber\\
&=&\int d{\bf x}_2 B(t|{\bf x}_2,t_2)B({\bf x}_2,t_2|{\bf
x}_1,t_1).
\end{eqnarray}
In the first line we inserted a $\delta$-function at time $t_2$
between times $t_1$ and $t$, and the second line is a consequence
of Markovian property. One may see this relationship is analogous
to the Chapman-Kolmogorov equation~(\ref{ChapmanKolmogorov}), and
a forward equation for $B({\bf x}_2,t_2|{\bf x}_1,t_1)$ can be
easily derived. On the other hand, the probability distribution
function of the time-reversed Eq.~(\ref{ReversedFPoperator}) at
time $s_1$$=$$t$$-$$t_1$ can be constructed by the distribution
function at earlier time $s_2$$=$$t$$-$$t_2$ given the transition
probability $p_{\rm R}$,
\begin{eqnarray}
\label{ChapmanEqTRerseForward} p({\tilde {\bf x}}_1,s_1)=\int
p_{\rm R}({\tilde {\bf x}}_1,s_1|{\tilde {\bf x}}_2,s_2)p({\tilde
{\bf x}}_2,s_2)d{\tilde {\bf x}}_2.
\end{eqnarray}
On the basis of Eq.~(\ref{decompositionTR}) and a comparison
between Eqs.~(\ref{ChapmanEqBackward})
and~(\ref{ChapmanEqTRerseForward}), we obtain
\begin{eqnarray}
\label{GeneralizedDetailbalance} p_{\rm R}({\tilde {\bf
x}}_1,s_1|{\tilde {\bf x}}_2,s_2)f({\bf x}_2,t_2)=B({\bf
x}_2,t_2|{\bf x}_1,t_1)f({\bf x}_1,t_1).
\end{eqnarray}
Here we used symbol $B()$ in Eq.~(\ref{evoleqofdynamicobsGS})
rather than $b()$ to indicate the generality of this identity. For
a time-reversible homogeneous stochastic system that was mentioned
previously, if we choose ${\bf S}={\bf S}^{\rm irr}(f)$ and
$f=f^{\rm eq}({\bf x})$, both the transition probability $p_{\rm
R}({\bf x},t|{\bf x}',t')$ ($t>t'$) of the time-reversed
system~(\ref{ReversedFPoperator}) and $B({\bf x},t|{\bf x}',t')$
defined here are identical with the transition probability
$\rho({\bf x},t|{\bf x}',t')$ in Eq.~(\ref{FPoperator}). Under
this consideration, the above identity is just the principle of
detail balance written in terms of conditional
probabilities~\cite{Gardiner,Risken}. An analogous expression has
been obtained earlier in Ref.~\cite{Chetrite08CMP} [Eq. (7.15)
therein] and was called generalized detailed balance relation. We
may point out that, compared with previous one the validity of
Eq.~(\ref{GeneralizedDetailbalance}) is larger.

Now we consider an ensemble average of a $(k+1)$-point function
over the time-reversed system~(\ref{ReversedFPoperator}),
\begin{eqnarray}
\label{ensembleaveragekpointfunction} &&\langle{G}[{{\textbf
x}}(s_0),\cdots,{{\bf x}}(s_k)]\rangle_{\rm R}= \int p_{\rm
R}({\tilde {\bf x}}_0,s_0|{\tilde {\bf x}}_{1},s_{1})\cdots p_{\rm
R}({\tilde {\bf x}}_{k-1},s_{k-1}|{\tilde {\bf x}}_k,s_k)f( {\bf
x}_k,t) {G}({{\bf x}}_0,\cdots,{ {\bf x}}_k)\prod_0^k{{d\tilde
{\bf x}}}_i,
\end{eqnarray}
where $s_k$$=$$t$$-$$t_k$,
$t$=$s_0$$>$$s_1$$>$$\cdots$$>$$s_k$$=$$0$, and we chose the
initial distribution $p(\tilde{\bf
x},0)$$=$$f(\varepsilon\tilde{\bf x},t)$. Employing
Eq.~(\ref{GeneralizedDetailbalance}) repeatedly, the right hand
side of the above equation becomes
\begin{eqnarray}
\int B({\bf x}_{k},t_{k}| {\bf x}_{k-1},t_{k-1})\cdots B({\bf
x}_1,t_1|{\bf x}_{0},t_{0})f({\bf x}_0,t_0) G({{\bf x}}_0,\cdots,{
{\bf x}}_k)\prod_0^k{{d {\bf x}}}_i,
\end{eqnarray}
Remarkably, letting $k$$\to$$\infty$ the function $G$ becomes a
functional ${\cal G}$ over the space of all stochastic
trajectories. We then obtain a very general identity
\begin{eqnarray}
\langle{\tilde{\cal G}}\rangle_{\rm R}= \langle {\cal
G}e^{-\int_0^t{\cal J}[f,{\bf S}]({\bf
x}(\tau),\tau)d\tau}\rangle, \label{GCrooks}
\end{eqnarray}
where ${\tilde {\cal G}}[\{\tilde{\bf x}(s)\}]={\cal G}[\{
\varepsilon{\tilde{\bf x}}(t-s)\}]$. Obviously, choosing ${\cal
G}$ to be a terminal function $B({\bf x}(t))$, we obtain the
GIFT~(\ref{GIFT}). Another important choice of the functional is
\begin{eqnarray}
\label{Crooksfunctional}  \delta(h-{\cal E}_t[\{{\bf
x}(\tau)\}])=\delta(h-\int_0^t{\cal J}[f,{\bf S}]({\bf
x}(\tau),\tau)d\tau).
\end{eqnarray}
Employing Eqs.~(\ref{reverseddrifts})
and~(\ref{reverseddiffcoeff}), it is easy to prove that the
overall relative entropy production functional ${\cal E}_t[\{{\bf
x}(\tau)\}]$ has the following property,
\begin{eqnarray}
\label{reversedentropybalancefunctional} &&{\tilde{\cal
E}}_t[\{\tilde{{\bf x}}(s)\}]= \int_0^t{\cal J}[f,{\bf
S}](\varepsilon{\tilde {\bf x}}(t-s),s)ds \nonumber\\
&=&-\{-\ln \frac{f({\varepsilon\tilde{\bf
x}}(t),0)}{f({\varepsilon\tilde{\bf x}}(0),t)}+({\rm S})\int_0^t
[2 \tilde{\hat{A}}_i^{\rm irr}({\tilde{{\bf
B}}}^{-1})_{ij}(\tilde{v}_j-{\tilde{A}}_j^{\rm
rev})-\partial_{{\tilde{x}}_i} {\tilde{A}}_i^{\rm rev}](
{\tilde{\bf x}}(s),s)ds\},
\end{eqnarray}
where $\tilde{v}_j=d\tilde{x}_j/ds$. Recalling the initial
distribution of the time-reversed process that was defined in
Eq.~(\ref{ensembleaveragekpointfunction}), the right hand side of
the second line is just the minus of the overall relative entropy
production functional ${\cal E}^{\rm R}_t[\{{\tilde{\bf x}}(s)\}]$
in the time-reversed system. This observation could be derived by
the involutive property of time reversal as
well~\cite{Chetrite08CMP}. Substituting the
functional~(\ref{Crooksfunctional}) into Eq.~(\ref{GCrooks}), we
obtain the transient DFT~\cite{Crooks00,Crooks99}
\begin{eqnarray}
P_{\rm R}(-h)=P(h)e^{-h}, \label{DFT}
\end{eqnarray}
where $P_{\rm R}(h)$ is the probability density for the stochastic
variable ${\cal E}^{\rm R}_t$$=$$h$ achieved from the reversed
process~(\ref{ReversedFPoperator}) with the specific initial
distribution mentioned above, and $P(h)$ is the probability
density for ${\cal E}_t$$=$$h$ achieved from the forward
process~(\ref{FPoperator}).

\section{Conclusion}\label{Conclusion}
In this work, we have tried to unify the derivations of the linear
response theory and the transient fluctuation theorems using the
perturbed Kolmogorov backward equations from a backward point of
view. The motivation of this reinvestigation of the linear
response theory is that conventional approach of the theory is
based on the forward Fokker-Planck equation and time-dependent
perturbation, which is not used in the FTs evaluations. Our
results show that, a derivation using the backward equation could
be very simple and flexible even if unperturbed system is
non-stationary. Importantly, this study also reminds us that the
time-invariable integral identity we found previously is the
generalization of the well-known Chapman-Kolmogorov equation. One
may notice that our evaluations heavily depend on the path
integral representation of the perturbed Kolmogorov backward
equations. Only in this representation, the physical relevances of
these partial differential equations appear explicitly. This
situation is very analogous to the relationship between the
Schr\"{o}dinger equation and Feynman path integral in quantum
physics. Hence one might criticize that these perturbed backward
equations are unnecessary because all above results could be
evaluated by direct path integral approach. This point is of
course correct in principle. However, as mentioned at the very
beginning, such a ``bottom-up" idea needs the known time reversal
or splitting of the drift vector. Except for very simple or
intuitive cases, e.g., those considered in
Sec.~\ref{priorsplitting}, finding a meaningful time reversal or
splitting is not trivial task. It would be desirable if there are
some rules or guides for this task. We think that these perturbed
backward equations satisfy this demand; see Eq.~(\ref{splitting}).
This is also logical. After all, the FTs are identities of
ensemble statistic properties of stochastic processes. In a word,
the roles played by these perturbed backward equations and their
path integral representations are complementary in the study of
the FTs. Considering that the generalized Chapman-Kolmogorov
equation is the cornerstone of this work, which intrinsically
arises from the Markovian characteristic of diffusion processes,
we believe that the evaluations and results developed here should
be also available to other Markovian stochastic processes, e.g.
general discrete jump processes with continuous~\cite{LiuF2} or
discrete time.
\\

{\noindent This work was supported in part by Tsinghua Basic
Research Foundation and by the National Science Foundation of
China under Grants No. 10547002 and No. 10704045. }

\appendix
\section{Limited Girsanov formula for degenerated diffusion matrix}\label{extendedCMGformula}
As that shown in Eq.~(\ref{CMGformula}), the Girsanov formula
requires the diffusion matrix to be positive definite. This point
may be better appreciated by first writing out the probability
density of a stochastic trajectory $\{{\bf x}(\tau)\}$ in the
stochastic system~(\ref{SDE})~\cite{ZinnJustin}:
\begin{eqnarray}
{\cal P}[\{{\bf
x}(\tau)\}]=[\prod_{k>M}^N\delta\left(\dot{x}_k-A_k \right)]\int
\prod_{1}^{M}{\cal D}[\eta_i]\exp[-\frac{1}{2}\int_0^t
\eta_i\eta_i ds]\prod_{i=1}^{M}\delta[ \dot{x}_i-A_i- ({\bf
D}^{\frac{1}{2}} )_{il}\eta_l], \label{pathprob}
\end{eqnarray}
where $\eta_i={dW_i}/{ds}$ is standard white noises, and
$\delta$-functions should be understood a product of a sequence of
terms on all times between $0$ and $t$. The expression in the
first square brackets on the right hand side indicates that noises
only act on the first $M$ coordinates. Assuming another stochastic
system (denoted by prime) has a different drift vector ${\bf
A}'$$=$${\bf A}$$+$${\bf a}$. Obviously, if there is any nonzero
component $a_k$ ($k$$>$$M$), the $\delta$ functions in the first
square brackets makes the ratio of the two probability densities
of the same trajectory in these two systems meaningless. In
physics this means we never observe the same trajectory in these
two systems. We met such a situation in the discussion of the FDTs
of the underdamped Brownian motion with the general
perturbations~\cite{comment2}; see sec.~\ref{FDRunderdampedBM}. On
the contrary, if nonzero components of ${\bf a}$ are restricted to
first $M$, namely, $a_k$$=$$0$ ($k$$>$$M$), the ratio or
Radon-Nikodym derivative of these two probability densities can be
always established and is
\begin{eqnarray}
\label{LGirsanovformula} {{\cal P}'[\{{\bf x}(\tau)\}]} = {{\cal
P}[\{{\bf x}(\tau)\}]}e^{-\int_{t_0}^t  \bar{\cal R}[{\bf
a}](\tau,{\bf x}(\tau))d\tau},
\end{eqnarray}
where the integrand is the same with Eq.~(\ref{CMGformula}) except
that the diffusion matrix ${\bf B}$ there is replaced by the
positive submatrix ${\bf D}$ and the summations are restricted to
first $M$ component. We call Eq.~(\ref{LGirsanovformula}) limited
Girsanov formula to distinguish with the standard one. We may
conveniently rewrite this limited formula into the standard one by
formally defining the inverse of the diffusion matrix ${\bf B}$;
see Eq.~(\ref{FormallyinversedB}) if we bear in mind the
application condition.


\begin{thebibliography}{99}
\bibitem{comment2}
This may be a reason why there are not FDTs analogous
Eq.~(\ref{overdampMaseFDR})~\cite{Baiesi} in the underdamped
Brownian motion with general perturbation depending on spatial and
momentum coordiantes simultaneously.
\end{thebibliography}
\end{document}